# Klein tunneling of the electronic states in the gate voltage modulated skyrmion crystal

Jianhua Gong[*] and Rui Zhu[†]

*School of Physics and Optoelectronics, South China University of Technology,*

*Guangzhou 510641, People's Republic of China*

**Abstract**

As a result of the Hund's coupling, the band structure of the conducting electrons in the skyrmion crystal (SkX) shares similar topological properties with that of graphene, such as its cone-like shape, nonzero band Chern number, edge states, and etc. In this work, we rigorously demonstrate that the Klein tunneling phenomena is also shared by these two. We use the Green's function technique and calculated the transmission probability of the electrons tunneling through an electrostatic barrier in the SkX expressed by the double exchange model. Numerical results of the SkX reproduced the Dirac model obtained by linear fitting the two-dimensional band structure of the SkX.

## I. Introduction

The magnetic skyrmion is the two-dimensional (2D) spin vortex texture, with the center spin pointing downward and the edge spin pointing upward. Experiment found that in most materials the skyrmions periodically form a lattice structure, so called the skyrmion crystal (SkX)[1,2,3,4,5]. In conducting SkX materials, as a result of the strong Hund's coupling, spin of the passing electrons tends to align to the direction of the local magnetization. This gives rise to a Peierls-phase-like site-dependent hopping integral and hence a reassembled band structure of the electrons[1,6,7]. The topologically nontrivial cone-like band structure is shared between the SkX and the multi-sublattice Dirac-Weyl materials such as graphene, Lieb lattice, Dice lattice, and etc. Similar transport properties originally hosted by the Dirac family have been found in the SkX such as the quantized Hall conductivity[7] and the topological phase transition driven by the next-nearest-neighbor hopping[8]. Naturally one wonders if we could go further to observe the most unconventional behavior of graphene -- the Klein tunneling.

The difficulty lies in the technique. The Green's function approach bridging the tight-binding (TB) model and the effective Dirac equation has been proposed in the tunneling process of monolayer graphene[9]. By considering each skyrmion as a giant unit cell, we extend the TB calculation in graphene to the double-exchange model in the SkX. Surface Green's function of the leads is calculated by the decimation method.

With the technique found, one motivation of the work is fulfilled, which is to pave way for a general numerical treatment of the transport property of multi-freedom crystal structures taking into account the background spin field and a complex unit cell. Recently, investigations in various systems have found that the transmission probability resembles as long as the energy dispersion has identical form regardless of the particular form of the Hamiltonian and its eigenvectors[9]. This encourages one to wonder whether the transport property of a complex Hamiltonian resembles the simple asymptotic Hamiltonian obtained by linear or parabolic band fitting. If so, one can predict the transport behavior of a complex system by its band structure in the low-energy regime if a rigorous calculation is beyond the current technique. This is the second motivation of the present work.

---

[*] Electronic address: 1208356262@qq.com
[†] Corresponding author. Electronic address: rzhu@scut.edu.cn

In this paper, we theoretically analyze the transport properties of the SkX by the Non-equilibrium Green's function (NEGF) and the Dirac model obtained from the band structure of the SkX. Strong consistency between the two theories is found in the low-energy regime. The NEGF approach developed in this work is not limited to low-energy excitation and linear dispersion restriction. It can be applied to wider conditions and provide more accurate results.

## II. Model and formalism

### A. Double exchange model in the skyrmion crystal

In order to simulate the transport properties of the SkX, we use the double exchange model to describe the free electron system coupled with the background spin texture[7]

$$H = \sum_i V_i c_i^\dagger c_i + t \sum_{i,j} c_i^\dagger c_j - J \sum_i \mathbf{n}_i c_i^\dagger \sigma c_i, \tag{1}$$

where $c_i = (c_{i\uparrow}, c_{i\downarrow})^T$ is the two-component annihilation operator at the $i$ site and $c_i^\dagger$ is its creation counterpart. $V_i$ corresponds to the on-site potential at site $i$ in the central region (see Fig. 1). $t$ is the hopping integral between nearest-neighbor sites and $J$ is Hund's coupling strength between the electron spin and the background spin texture. $\mathbf{n}_i = (\cos\Phi_i \sin\Theta_i, \sin\Phi_i \sin\Theta_i, \cos\Theta_i)$ is the spin configuration in spherical coordinates. $\boldsymbol{\sigma} = (\sigma_x, \sigma_y, \sigma_z)$ indicates the vector of Pauli matrices.

In the strong-coupling limit, i.e., $J \gg t$, the spin of electron on site is forced to align parallel to the background spin texture. The effective hopping strength is determined by the spin overlap between neighboring sites $i$ and $j$, i.e., the effective hopping integral between nearest-neighbor sites is obtained by $\langle \chi_i | \chi_j \rangle$. The wave function $|\chi(\mathbf{r})\rangle$ of electron is the spin eigenstate of $n(\mathbf{r})$ at $\mathbf{r}$. Using spherical coordinates in the spin space,

$$|\chi(\mathbf{r})\rangle = \left(\cos\frac{\Theta(\mathbf{r})}{2}, e^{i\Phi(\mathbf{r})} \sin\frac{\Theta(\mathbf{r})}{2}\right)^T, \tag{2}$$

and

$$t_{eff} = t \langle \chi_i | \chi_j \rangle = \cos\frac{\Theta_i}{2}\cos\frac{\Theta_j}{2} + \sin\frac{\Theta_i}{2}\sin\frac{\Theta_j}{2}e^{-i(\Phi_i-\Phi_j)}. \quad (3)$$

In this limit, the effective Hamiltonian reduces the degree of freedom of the spin[7]. Then,

$$H = \sum_i V_i d_i^\dagger d_i + \left( \sum_{i,j} t_{eff} d_i^\dagger d_j + \text{H.c.} \right). \quad (4)$$

Here, $d_i^\dagger$ is the spinless creation (annihilation) operator. $\theta_i$ and $\phi_i$ represent the polar and azimuthal angles on the $i$ site, respectively. The skyrmion profile is well assumed as

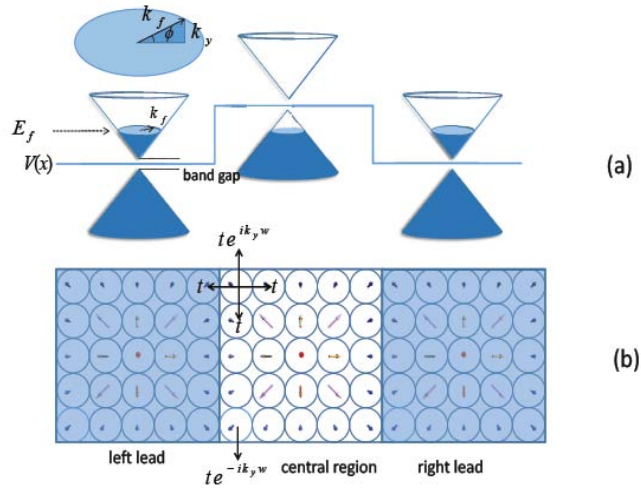

FIG1: (a) Schematic image of the $npn$ junction of the Dirac theory with an energy gap. The blue filled areas indicate occupied states. The Fermi level lies in the conduction band in the left and right leads and in the valence band inside the barrier. The small picture above is the top view of the band showing the incident angle and the Fermi ring. (b) Schematic image of the $npn$ junction in the SkX consistent with that of panel (a). Shaded regions indicate the leads and the transparent region indicates the central barrier. The square made up of $5\times 5$ circles indicates the minimum unit that simulates the bulk SkX with $W$ the transverse periodicity. The arrows describe the background spin of the SkX.

$\Theta(r) = \pi(1 - r/\lambda)$ for $r < \lambda$ and $\Theta(r) = 0$ for $r > \lambda$, where $\lambda$ is radius of the skyrmion and $(r, \varphi)$ are the polar coordinates in the real space counting from the center of each skyrmion. When the skyrmion whirls in the pattern of $\Phi(\varphi) = \xi\varphi + \gamma$, the topological number is explicitly expressed by $Q = m\xi$ and $\gamma$ determines the helicity of the skyrmion. $\gamma=0(\pi/2)$ correspond to the Néel type and Bloch type skyrmions, respectively. The vorticity $m = \pm 1$ correspond to skyrmions and antiskyrmions, respectively.

We consider each skyrmion unit cell contains $5\times 5 = 25$ atoms, setting the radius $\lambda = 2.5a$.

$a$ is the atomic lattice constant. We construct an $npn$ junction shown in Fig. 1 (a) or an $nn'n$ junction (in which the barrier is low and the Fermi level lies in the conduction band inside as well as outside the barrier) in the SkX. Due to the translation invariance in the transverse direction, we impose the Bloch theorem along the transverse direction with periodicity $W = 2\lambda = 5a$. Applying the boundary condition of period $W$, we modify the hopping strength between neighboring unit cells in the transverse direction by a Bloch phase factor $e^{\pm ik_y W}$ shown in Fig.1(b). The whole system can be described by a one-dimensional chain. We get the relationship between the incident angle $\phi$ and $k_y$ on the Fermi ring, as shown in the upper part of Fig.1 (a). It's worth noting that the shape of the Fermi ring is different for different Fermi energy. Fig. 6 shows the Fermi ring of the lowest two bands of the SkX. If the selected Fermi energy is close to the peak of the Dirac cone, the Fermi ring here is basically a circle. We can use $k_y = k_f \sin\phi$ to calculate the angular dependence of the transmission probability.

The central region is finite in the transport direction. We perform the Fourier transform of the Hamiltonian along the $y$ direction. Then, the Fourier transform is written as $d_{mi} = \sum_{k_y} d_{k_y} e^{ik_y Y_{mi}}$. The index "$m$" represents the position of the unit cell along the x direction. By diagonalization of the Hamiltonian in the ribbon space, the energy spectrum as a function of the momentum $k_y$ is shown in Fig. 7 (a) and (c).

**B. Non-equilibrium Green's function**

The Non-equilibrium Green's function (NEGF) formalism provides a simple and effective method to describe the quantum transport at nanoscale[10]. Even so, there are so many atoms in the system that it makes our calculations difficult. By applying the Bloch's theorem in the transverse direction[9,11], the system can be reduced to a one-dimensional chain. For a certain band energy $E$ and transverse wavevector $k_y$, we can numerically compute the retarded and advanced Green's function as,

$$G^r_{k_y}(E) = \left(G^a_{k_y}(E)\right)^\dagger = \left((E+i\eta)I - H_{C,k_y} - \Sigma^r_{L,k_y} - \Sigma^r_{R,k_y}\right)^{-1}. \quad (5)$$

Here, $\Sigma^r_L = H_{CL} g^L_{00} H_{LC}$ ($\Sigma^r_R = H_{CR} g^L_{00} H_{RC}$) is the retarded self-energy of the left (right) lead. The surface Green's function $g^L_{00}$ ($g^R_{00}$) of the leads can be solved by the decimation method[12]. $H_C$ indicates the Hamiltonian for the central region. $H_{CL}/H_{CR}$ is the Hamiltonian connecting the central region and the left/right leads and $H_{LC}/H_{RC}$ is its Hermitian conjugate. $\eta$ and $I$ represent an infinitely small positive number and the identity matrix, respectively.

We calculate the transmission probability for an incident electron by the Fisher-Lee relation[13]

$$T(E, k_y) = \text{Tr}(\Gamma_L G^r \Gamma_R G^a), \tag{6}$$

where the spectral functions was computed by $\Gamma_{L/R} = i\left(\Sigma^r_{L/R} - \Sigma^a_{L/R}\right)$. When the Fermi energy $E_f$ of the incident electron is determined, the correspondence between Bloch momentum $k_y$ and the incident angle $\phi$ is determined from the fermi ring.

According to the Landauer-Büttiker formula, tunneling conductance of the electrons as a function of the Fermi energy can be expressed as

$$G(E_f) = \frac{2e^2}{h} \frac{W}{2\pi} \int T(E_f, k_y) dk_y, \tag{7}$$

where $G_0 = e^2 L/(h\pi)$.

It is time consuming to directly invert the large matrix at the right hand side of Eq (5) when the central region is large. We adopt the recursive Green's function to avoid this problem[11,14,15], which is based on the Dyson equation. The central region is divided into slices, which only couple with neighboring slices as shown in Fig. 2 (a). Each slice is a single unit cell in the transport direction and extends to infinity in the transverse direction.

Numerically, we use a recursive process from left to right, or vice versa. At the first step, the Green's function

$$G_{11}^{\rightarrow(1)}(k_y) = (E - H_{11}(k_y) - \Sigma_L(k_y))^{-1}. \tag{8}$$

The $G_{11}^{\rightarrow(1)}$ means that the first slice is attached to the left lead shown in Fig. 2 (b). We repeat this procedure until the last slice $N$,

$$G_{ii}^{\rightarrow(i)}(k_y) = \left(E - H_{ii}(k_y) - \Sigma_R(k_y)\delta_{iN} - H_{i,i-1} G_{i-1,i-1}^{\rightarrow(i-1)}(k_y) H_{i-1,i}\right)^{-1}, \tag{9}$$

$$G_{1i}^{\rightarrow(i)}(k_y) = G_{1,i-1}^{\rightarrow(i-1)}(k_y) H_{i-1,i}(k_y) G_{ii}^{\rightarrow(i)}(k_y). \tag{10}$$

When the first slice is attached to the left lead, a new lead surface is created, which is then attached to the second slice by a new surface Green's function $G_{11}^{\rightarrow(1)}$. The next step is that a next new surface is attached to the third slice by a next new surface Green's function $G_{22}^{\rightarrow(2)}$ until it is connected to the right lead. Through this recursive process, the entire central system has been taken into account. See Reference [11,14,15] for more details. Finally, the transport coefficient of the system can be obtained by

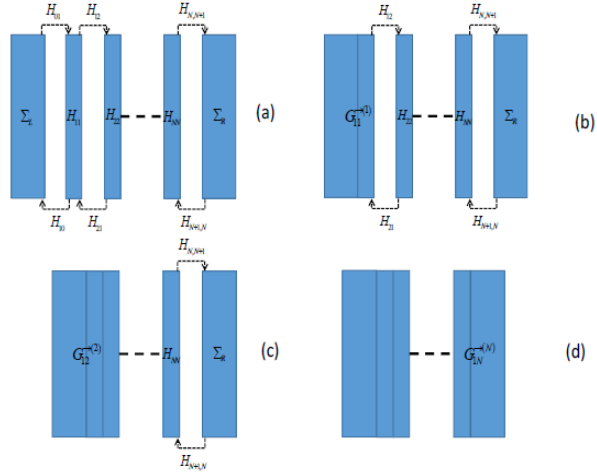

FIG2: The process of recursive Green's function. (a) The recursive process has not yet been carried out. (b) The first step in the recursive process. (c) The second step of the recursive process. (d) The last step of the recursive process.

$$T(E, k_y) = \mathrm{Tr}(\Gamma_L G_{1N}^r \Gamma_R G_{1N}^a). \tag{11}$$

### C. Dirac theory

In the strong coupling ($J \gg t$) limit, the structure of the lowest and second-lowest band has the properties of the Dirac cone. We can describe this electronic structure of massive Dirac fermions by the Dirac theory[7]

$$H = \hbar v(-k_x \sigma_x + k_y \sigma_y) + m\sigma_z, \tag{12}$$

where, $v$ is the Fermi velocity and the mass $m$ is determined by the gap $\Delta$ as $m = \Delta/2$.

The eigenvalue of Eq (12) is $E = \lambda\sqrt{\hbar^2 v^2 k^2 + m^2}$, $\lambda = \pm 1$ correspond to the conduction and valence bands, respectively. One assumes that the wave function of Eq (12) is $\psi_{\lambda,k} = \frac{1}{2}(\mu_\lambda, \upsilon_\lambda)^\mathrm{T} e^{i(k_x x + k_y y)}$. Putting the hypothetical wave function into Eq (12), we get

$$\psi_{\lambda,k} = \frac{1}{\sqrt{2}} \begin{pmatrix} \sqrt{1 + \frac{\lambda m}{\sqrt{m^2 + \hbar^2 v^2 k^2}}} \\ -\lambda \sqrt{1 - \frac{\lambda m}{\sqrt{m^2 + \hbar^2 v^2 k^2}}} e^{-i\phi_k} \end{pmatrix} e^{i(k_x x + k_y y)}. \tag{13}$$

When the electron is incident into the square barrier $V(x) = V_0$ ($0 \leq x \leq D$) shown in Fig.1 (a). Then the wave function in three different regions is

$$\psi_i = \frac{1}{\sqrt{2}} \begin{pmatrix} \alpha \\ -\lambda\gamma e^{-i\phi_k} \end{pmatrix} e^{i(k_x x + k_y y)} + \frac{r}{\sqrt{2}} \begin{pmatrix} \alpha \\ -\lambda\gamma e^{-i(\pi - \phi_k)} \end{pmatrix} e^{i(-k_x x + k_y y)}, \tag{14}$$

$$\psi_{ii} = \frac{a}{\sqrt{2}} \begin{pmatrix} \beta \\ -\lambda'\eta e^{-i\theta_q} \end{pmatrix} e^{i(q_x x + k_y y)} + \frac{b}{\sqrt{2}} \begin{pmatrix} \beta \\ -\lambda'\eta e^{-i(\pi - \theta_q)} \end{pmatrix} e^{i(-q_x x + k_y y)}, \tag{15}$$

$$\psi_{iii} = \frac{t}{\sqrt{2}} \begin{pmatrix} \alpha \\ -\lambda\gamma e^{-i\phi_k} \end{pmatrix} e^{i(k_x x + k_y y)}. \tag{16}$$

Where $\alpha = \sqrt{1 + \frac{\lambda m}{\sqrt{m^2 + \hbar^2 v^2 (k_x^2 + k_y^2)}}}$, $\gamma = \sqrt{1 - \frac{\lambda m}{\sqrt{m^2 + \hbar^2 v^2 (k_x^2 + k_y^2)}}}$, $\beta = \sqrt{1 + \frac{\lambda' m}{\sqrt{m^2 + \hbar^2 v^2 (q_x^2 + k_y^2)}}}$, $\eta = \sqrt{1 - \frac{\lambda' m}{\sqrt{m^2 + \hbar^2 v^2 (q_x^2 + k_y^2)}}}$.

According to the continuity condition of the wave function, which is that the wave function is continuous at $x = 0$ and $x = D$. Then,

$$\alpha + r\alpha = a\beta + b\beta, \tag{17}$$

$$-\lambda\gamma e^{-i\phi_k} + r\lambda\gamma e^{i\phi_k} = -a\lambda'\eta e^{-i\theta_q} + b\lambda'\eta e^{i\theta_q}, \tag{18}$$

$$a\beta e^{iq_x D} + b\beta e^{-iq_x D} = t\alpha e^{ik_x D}, \tag{19}$$

$$-a\lambda'\eta e^{-i\theta_q} e^{iq_x D} + b\lambda'\eta e^{i\theta_q} e^{-iq_x D} = -t\lambda\gamma e^{-i\psi_k} e^{ik_x D}. \tag{20}$$

After some algebra, we can get transmission probability

$$T = |t|^2 = \frac{\cos(\theta_q)^2 \cos(\phi_k)^2}{\sin(q_x D)^2 \left(\frac{N}{\lambda\lambda'\eta} - 2\sin(\phi_k)\sin(\theta_q)\right)^2 + (\cos(\phi_k)\cos(\theta_q)\cos(q_x D))^2}. \tag{21}$$

In the transmission probability $\frac{N}{\lambda\lambda'\eta} = 2\frac{\lambda'}{\lambda}\left(\frac{|V_0 - E|E - \lambda\lambda' m^2}{\hbar^2 v^2 kq}\right)$, $V_0 - E = \sqrt{m^2 + \hbar^2 v^2 q^2}$, $E = \sqrt{m^2 + \hbar^2 v^2 k^2}$, $q = \sqrt{k_y^2 + q_x^2}$, $k = \sqrt{k_y^2 + k_x^2}$. The injection and refraction angles of the electron are $\phi_k = \arctan(\frac{k_y}{k_x})$ and $\theta_q = \arctan(\frac{k_y}{q_x})$, respectively.

For normal incidence, i.e. $\phi_k = 0$,

$$T(\phi_k = 0) = \frac{1}{1 + \sin(q_x D)^2 \left( \left( \frac{N}{\lambda \lambda' \eta} \right)^2 - 1 \right)}. \tag{22}$$

Because $\frac{N}{\lambda \lambda' \eta} = 2\frac{\lambda'}{\lambda} \left( \frac{|V_0 - E|E - \lambda \lambda' m^2}{\hbar^2 v^2 kq} \right) \neq \pm 1$, perfect transmission occurs when the condition $q_x D = n\pi (n = 0, 1, 2, 3 \cdots)$ is satisfied.

**III. Results and discussions**

**A. Klein tunneling in the strong coupling limit**

In this section, we compare the tunneling results between the NEGF and the Dirac theory in the case of $J \gg t$. In the NEGF treatment, we arbitrarily selected three Fermi energies of the incident electron: $E_f = -3.1t$, $-3.08t$, and $-3.12t$ corresponding to $E_f = 0.0825t$, $0.1025t$, and $0.0625t$ in Dirac theory, the latter of which is counted from the charge neutrality point. The width of the barrier is set to be $D = 100$ nm. In order to exactly match the barrier width, we set the lattice constant $a = 1$ nm in the double exchange model. In this case, we can estimate the Fermi velocity from the SkX band to compare with the asymptotic Dirac theory by calculating the slope of the dispersion. Fig. 3 show the transmission probability obtained from the NEGF and the Dirac theory, the two of which demonstrate close accordance. Fig. 3 (b) and the black curves in Fig. 3 (c) demonstrate perfect transmission at normal incidence, which strongly confirms the occurrence of the Klein tunneling in the gate voltage moduated SkX. Also, the NEGF and Dirac theory consistently demonstrate multiple transmission peaks at other incident angles, which results from coaction of resonance and Klein tunneling. We should note that there is still an angle difference of about $5°$, which reflects the difference between the two methods, which was addressed in other systems[9].

Fig 3 (b) demonstrates a difference between the two theories: there is one less transmission peak in the transmission probability obtained from the Dirac theory than that from the NEGF calculation. The reason is that the Dirac theory describes the linear dispersion relationship concerning low energy excitation, which means that the Fermi ring in the energy space is circular. But a real Fermi ring of the electronic states in the SkX is not a circle, which means that the Fermi velocity and $q_x$ are not fixed. We can get this information from the band structure. Because the small difference in the incident angle obtained from the two theories, the missing transmission peak has gone beyond $90°$ and is therefore missing.

In the Dirac theory, it is predicted that the transmission peak will appear on $q_x D = n\pi$ at normal incidence. We keep the width $D$ of the barrier constant and change the height $V_0$ of the barrier to modify $q_x$ by $q_x = \sqrt{(V_0 - E)^2 - m^2/(\hbar^2 v^2)}$. Results of the calculation are shown

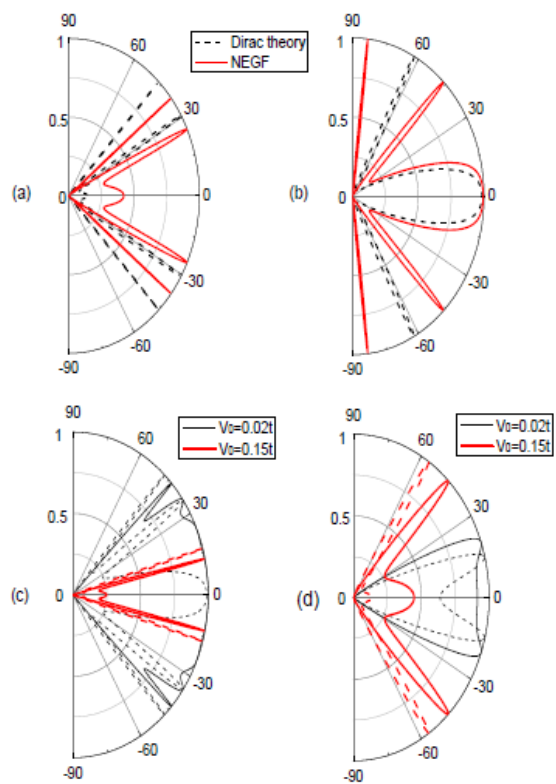

FIG3: Transmission probability $T(E_f, \phi)$ of the SkX obtained from the NEGF calculation and Eq. (21) of the Dirac theory. Solid and dashed lines are results of the NEGF and Dirac theory, respectively. In panels (a) and (b), $E_f = -3.1t$, $V_0 = 0.15t$ and $0.2t$, respectively. In panels (c) and (d), $E_f = -3.08t$ and $-3.12t$, respectively.

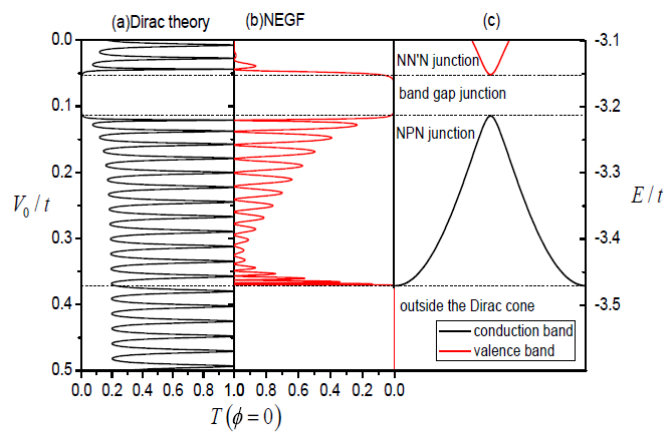

FIG4: Transmission probability as a function of barrier height $V_0$ at normal incidence. (a) and (b) are calculated from the Dirac theory and the NEGF, respectively. The Fermi energy of the incident electron is $E_f = -3.1t$ for NEGF. (c) represents the Fermi energy position corresponding to the height of the central barrier.

in Fig. 4 (a). Since $q_x$ is not involved in the NEGF calculation, we only need to modify $V_0$ in the Hamiltonian of the central region. The corresponding results are shown in Fig. 4 (b).

When the height of the barrier $V_0 \leq 0.0514t$, the incident electron travels from the conduction band at the left lead through the conduction band at the barrier and finally reaches the conduction band at the right lead, which is an $nn'n$ junction. When the barrier goes up to $0.0514t \leq V_0 \leq 0.1136t$, the Fermi energy of the incident electron is in the energy gap of the barrier region. Because neither the electron nor the hole state fills the gap, the incident electron lacks resonant level at the barrier and crossing of the electron is prevented. The transmission probability is zero disregarding whether we use the Dirac theory or the NEGF. As the barrier rises further, the Fermi energy locates in the valence band in the barrier region. The electron transport turns into hole transport in the valence band and finally turns back into electron transport in the conduction band in the right lead, giving rise to the Klein tunneling effect. This is an $npn$ junction. If we lifted the central barrier even higher, the Fermi energy exceeds the Dirac cone band range, no transmission occurs, which is the same as in the band gap.

In the case of low energy excitation, the transmission property obtained by Dirac theory is consistent with the NEGF. In the case of high energy excitation, nonlinear effect of the band is obvious and the Dirac theory is no longer applicable. Nevertheless the NEGF method can be applied at any energy.

**B. Transmission probability of finite Hund's coupling**

In this section, we analyze transmission probability in the case of finite Hund's coupling. In this case, influence of the electron spin subjected to the background spin texture is limited. We have to take into account the two spin directions of the electron. A transmission probability of $2$ means that electrons in both spin directions can cross the barrier completely. Because the range of the energy bands is different under different Hund's coupling strength $J$. As a result of the finite

Hund's coupling, the position of the energy bands changes a little in comparison with the case of the infinite Hund's coupling. To compare the two cases and to see a strong Klein tunneling effect, we tune the Fermi energy and the barrier height according to the band structure for different Hund's coupling strength to have the Fermi energy $0.02t$ higher than the conduction bottom in the leads and simultaneously $0.02t$ lower than the valence band top in the barrier. In this way, filling pattern of the electrons in the leads and that of the holes in the barrier is the same.

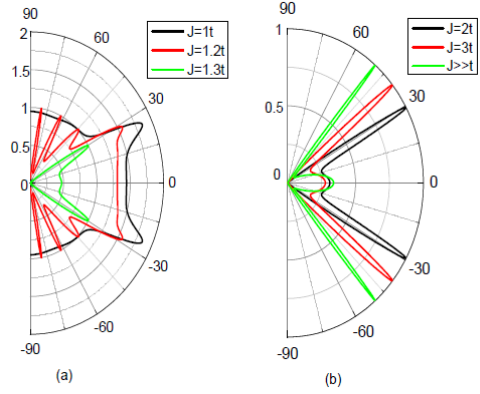

FIG. 5: Transmission probability of finite Hund's coupling. For $J = 1t$, $E_f = -4.1807t$, $V_0 = 0.067t$; for $J = 1.2t$, $E_f = -4.3769t$, $V_0 = 0.0698t$; for $J = 1.3t$, $E_f = -4.4752t$, $V_0 = 0.0711t$; for $J = 2t$, $E_f = -5.1662t$, $V_0 = 0.0777t$, for $J = 3t$, $E_f = -6.1583t$, $V_0 = 0.0834t$; for $J \gg t$, $E_f = -3.1314t$, $V_0 = 0.1022t$. Other parameters are the same as Fig. 3.

In Fig. 5 (a), when $J < 1.3t$, the transport coefficient exceeds 1 because of the electron spin is not aligned with the background SkX and the spin degree of freedom is taken into account. When $J \geq 1.3t$, the electron spin effect disappeared, which shows that the Hund's coupling strength is sufficient to restrain the direction of electron spin deflection. As the coupling strength increases, the shape of the transport probability $T$ gradually changes to that obtained in the strong coupling limit by comparison of Fig. 5 (b) and Fig. 3.

**C. Transmission probability and band**

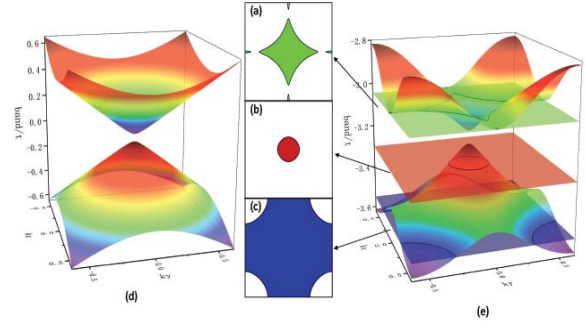

FIG6: (d) The bands described by the Dirac equation (12) and (e) the lowest two bands described by the effective tight-binding model of the SkX under the strong coupling condition $J \gg t$. The corresponding energies of Fermi cross profile (a), (b) and (c) are $E_f = -3.0438t$, $E_f = -3.3t$, $E_f = -3.55t$.

We analyze the relationship between the transmission probability and the band configuration by concerning the edge states. In this part, we mainly discuss the situation of $J \gg t$, where the Dirac cone of the SkX is easier to observe as shown in Fig. 6 (d). Gradually increasing the barrier height in the central region can change the band position of the central region where the Fermi energy is aligned. The Fermi energy of $E_f = -3.1t$ in the left lead aligns with the energy of $E = -3.3t$ in the valence band of the central region when the barrier height is $V_0 = 0.3t$, where the cross profile belongs to the Dirac cone energy shown in Fig. 6 (b). The boundary of the Dirac cone is shown in Fig. 6 (a), where results of the NEGF began to deviate from the Dirac theory. The outside of the Dirac cone cannot contribute to transport, the Fermi cross profile of which is shown in Fig. 6 (c), where total reflection occurs at all incident angles.

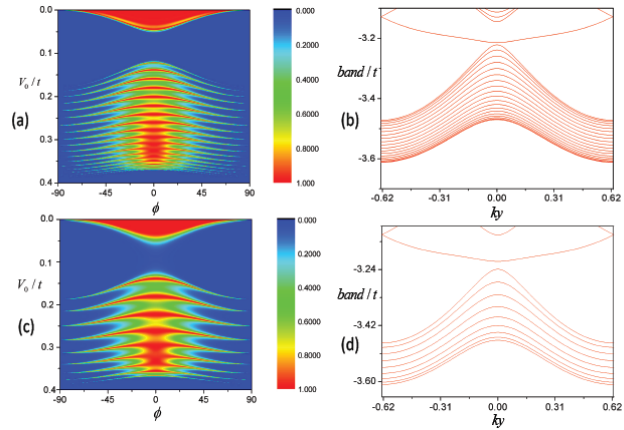

FIG7: (a) and (c), Transmission probability as a function of the incident angle $\phi$ and the barrier height $V_0$. (b) and (d), band structure of the central nanoribbon (the central barrier region can be looked at as a nanoribbon along the transverse direction of the transport) with barrier widths

$D = 20$ skyrmions (one skyrmion constitutes one slice in the NEGF treatment) and $10$ skyrmions, respectively. The Fermi energy is set to be $E_f = -3.1t$.

Calculating transmission probability as a function of the incident angle $\phi$ and the barrier height $V_0$ at $E_f = -3.1t$ can capture three pieces of information. First, total reflection appears inside the band gap and outside of the Dirac cone in the central region. Second, the number of lines with a transmission coefficient of $1$ for the $npn$ junction is the same as the number of the valence band line of the nanoribbon geometry. Fig. 7 (b) shows that the valence band line of the nanoribbon geometry with a barrier width of $20$ slices is only $19$, because that the missing line is the edge state inside the gap. It is interesting that the line of the transmission coefficient of $1$ for the $npn$ junction is also $19$ as shown in Fig. 7 (a). The same phenomenon also occurs in the case of $D = 10$ slices shown in Fig. 7 (c) and (d). This demonstrates that tunneling of the electron occurs through the eigenstates of the middle barrier. Third, besides the number of the perfect transmission peaks, the sparse and dense positions of the two are basically the same. The relation between the transport property and the band structure of the material is further confirmed.

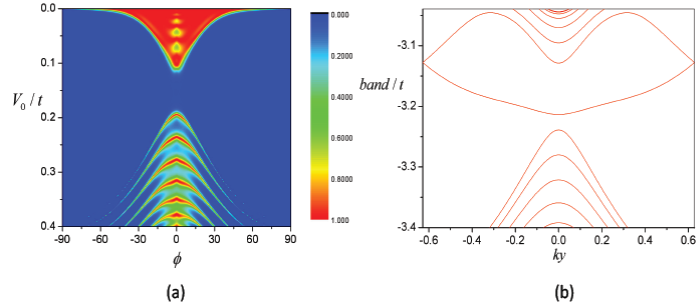

FIG8: (a) Transmission probability as a function of the incident angle $\phi$ and the barrier height $V_0$. (b) Energy bands of the nanoribbon in the central region. $D = 10$ slices and $E_f = -3.044t$.

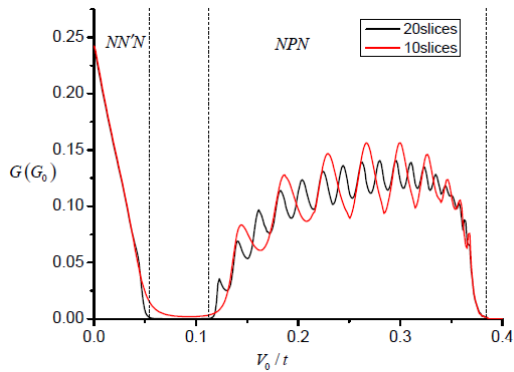

FIG9: Change in conductance with taking into account potential barrier. The Fermi energy of the incident electron is $E_f = -3.1t$. The red line and the black line correspond to $10$ slices and $20$ slices in the central region, respectively.

The same situation occurs with $nn'n$ junction. But it is difficult to observe how many lines with a transmission coefficient of $1$ there are, because the lines above are too dense to merge together as shown in Fig. 8 (a). The large red part in the upper half of the figure is the transmission probability of the $nn'n$ junction, in which the incident level aligned with one level in the conduction band of the central region. The bottom half of the figure is the $npn$ junction, in which the incident level aligned with one level in the valence band of the central region. It is again shown that the transmission characteristics coincide with the dispersion of the quasiparticle eigenstates of the central nanoribbon.

In the end, by calculating the conductance $G$ we find that the conductance of the $nn'n$ junction is close to a linear reduction corresponding to the linear reduction of the density of states of the conduction band as the barrier increases, which can be interpreted by the upper half red-color region in Fig. 8 (a). For the $npn$ junction, discrete levels in the valence band give rise to oscillation in the conductance, which can be interpreted by Fig. 8 (b).

## IV. Conclusions

In this work, we use the NEGF method and the Dirac theory to analyze the transport property of the electron states in the gate voltage modulated SkX. The infinite and finite Hund's coupling conditions are taken into account consistently. The NEGF method is applicable in a wider parameter range than the Dirac theory and by analyzing the Fermi ring the angular dependence of the transmission probability can be obtained. The Fermi velocity corresponding to a non-circular Fermi ring is not fixed, which leads to the difficulty of the Dirac theory to predict the electron transport in the SkX. The NEGF method avoids the calculation of Fermi velocity, which leads to more accurate numerical results. We also calculate the transmission probability by taking into account the finite Hund's coupling, which exceeds $1$ because the spin degree of freedom is released. As $J$ increases to $1.3t$, the maximum value of the transmission probability changes back to $1$, because the Hund's coupling is strong enough to restrain the electron spin along the back ground SkX. By comparing the contour of the transmission probability in the $\phi$-$V_0$ space and the band structure of the nanoribbon geometry in the central barrier region, we have reached a conclusion that the energy band of the central region affects the transmission probability. As the potential barrier increases, the conductance of the $nn'n$ junction decreases linearly for small barrier height and oscillation occurs in the $npn$ junction for large barrier hight, which can be interpreted by the eigenlevel configuration of the central nanoribbon.

## V. Acknowledgements

We acknowledge support by the National Natural Science Foundation of China (No. 11004063) and the Fundamental Research Funds for the Central Universities, SCUT (No. 2017ZD099).